# Semiconducting Metal Oxide Photonic Crystal Plasmonic Photocatalysts

Gillian Collins[1,2,3], Alex Lonergan[1], David McNulty[1], Colm Glynn[1], Darragh Buckley[1], Changyu Hu[1], and Colm O'Dwyer[1,2,3,4]

[1]*School of Chemistry, University College Cork, Cork, T12 YN60, Ireland*

[2]*Tyndall National Institute, Lee Maltings, Cork, T12 R5CP, Ireland*

[3]*AMBER@CRANN, Trinity College Dublin, Dublin 2, Ireland*

[4]*Environmental Research Institute, University College Cork, Lee Road, Cork T23 XE10, Ireland*

**Abstract**

Plasmonic photocatalysis has facilitated rapid progress in enhancing photocatalytic efficiency under visible light irradiation. Poor visible-light-responsive photocatalytic materials and low photocatalytic efficiency remain major challenges. Plasmonic metal-semiconductor heterostructures where both the metal and semiconductor are photosensitive are promising for light harvesting catalysis, as both components can absorb solar light. Efficiency of photon capture can be further improved by structuring the catalyst as a photonic crystal. Here we report the synthesis of photonic crystal plasmonic photocatalyst materials using Au nanoparticle-functionalized inverse opal (IO) photonic crystals. A catalyst prepared using a visible light responsive semiconductor ($V_2O_5$) displayed over an order of magnitude increase in reaction rate under green light excitation ($\lambda$=532 nm) compared to no illumination. The superior performance of Au-$V_2O_5$ IO was attributed to spectral overlap of the electronic band gap, localized surface plasmon resonance and incident light source. Comparing the photocatalytic performance of Au-$V_2O_5$ IO with a conventional Au-$TiO_2$ IO catalyst, where the semiconductor band gap is in the UV, revealed that optimal photocatalytic activity is observed under different illumination conditions depending on the nature of the semiconductor. For the Au-$TiO_2$ catalyst, despite coupling of the LSPR and excitation source at $\lambda$=532 nm, this was not as effective in enhancing photocatalytic activity compared to carrying out the reaction under broadband visible light, which is attributed to improved photon adsorption in the visible by the presence of a photonic band gap, and exploiting slow light in the photonic crystal to enhance photon absorption to create this synergistic type of photocatalyst.

**Introduction**

Plasmonic nanoparticles strongly absorb visible light due to their localised surface plasmon resonance resulting in photo generated electrons and holes which can be utilized for enhanced photocatalysts. Direct plasmonic photocatalysis describes chemical transformations which occur at the surface of the plasmonic NP under the excitation of the LSPR. [1] A variety plasmon induced chemical reactions have been reported such as oxidation[2], coupling reactions[3], $H_2$[4] and, S-S[5] bond dissociation, and others which have been subject to extensive reviews.[6] Another general class of plasmon-enhanced photocatalysts are based on metal-semiconductor heterostructures which have shown promise for photocatalytic conversion of light energy.[7-9] Incorporating plasmonic NPs into wide band gap semiconductors, such as Au-$TiO_2$ systems, extends light absorption of the semiconductor into the visible range as plasmon-generated hot electrons can transfer to the semiconductor thereby prolonging the lifetime of charge carriers.[7, 10] Electrons transferred to the semiconductor drive reduction processes and holes left in the metal NP can induce oxidative transformations enabling applications for water splitting,[11] $CO_2$ reduction[12], pollutant degradation[13], aerobic oxidation[14] and selective hydrogenation of cinnamaldehyde[15], as an example. Combining plasmonic NPs with narrow band gap (vis-NIR) semiconductors has been less studied in comparison, but these systems enable both the metal and the semiconductor support to be sensitized by visible light. When the plasmon resonance of the NPs overlaps with the electronic absorption of the semiconductor, incident light simultaneously excites the LSPR in the NPs and generates electron-hole pairs in the semiconductor giving rise to strong local field effects.[16] Charge transfer can occur from metal to semiconductor or vice-verse depending on the nature of the band alignment.[8] Super-resolution mapping of photogenerated electrons in Au-tipped CdS heterostructures verified the existence of two fundamentally distinct charge separation mechanisms in systems when both the metal and semiconductor are excited by the incident light source.[17] Therefore, through selection of the appropriate metal NP and semiconductor support, such heterostructures potentially enable the flow of electrons to be for optimized for a particular reaction under visible light.

In addition to a well-designed energy band structure to modulate charge-carrier generation and migration, the overall catalyst architecture plays an important role in the photocatalytic enhancement observed. In particular, the semiconductor superstructure influences the migration of excited electrons and enable longer charge carrier lifetimes.[18] At the semiconductor-solution interface, small dimensions that approach the depletion layer width of the semiconductor can deplete the material of majority carriers, but physically structuring the semiconductor to maximize ionic and electronic mobility can be very useful. The use of photonic crystals as catalyst support architectures has generated considerable interest for light harvesting catalytic applications as their unique optical and structural features to be exploited for enhanced performance. [19] Photonic crystals influence the propagation of light by their periodic variation in dielectric contrast .[20] One very useful property of photonic crystals is the phenomenon of slow group velocity photons, or 'slow' light.[21] At energies close to the photonic band gap, the group velocity of light is retarded, giving rise to the slow photon effect[22] which can increase the degree of light absorption and can been exploited for photocatalytic applications. Furthermore, trapping of light at standard group velocities by a photonic crystal that acts as a dielectric mirror may also contribute to enhanced photocatalysis. By combining plasmonic and photonic nanostructures, it is possible to tune the electronic absorption of the semiconductor bandgap with the LSPR of the metal NP and a photonic band gap of the semiconductor support. This synergy can be used to maximize the photon-to-electron efficiency for electron injection into the oxide conduction band or the metal NP surface, depending on the barrier height and alignment. Zhang *et al.*[23] assembled Au NPs on $TiO_2$ nanotube arrays where the LSPR wavelength matched the photonic band gap of the $TiO_2$ support, increasing hot electron injection and improved performance in photo electrochemical (PEC) water splitting. Other nanostructures effectively integrating the photonic band gap with the SPR include Au-$TiO_2$ bilayer nanorod-photonic crystal[24] for water PEC systems, thin shell Au/$TiO_2$ hollow nanospheres for decomposition of isopropanol to $CO_2$[25] and Au-ZnO photonic crystals for degradation of rhodamine blue[26]. The plasmonic effect of Au NPs deposited on visible light active semiconductors bismuth vanadate ($BiVO_4$)[27] and CdS-Au-$WO_3$[28] was significantly amplified due to strong coupling with the photonic Bragg resonance, resulting in high performance catalysts for photocatalytic water splitting. In addition to the unique optical properties of IO photonic crystals, the interconnected macroporous and microporous networks enables favourable mass transport of reactants in solutions[29]. The IO architecture remains electrically interconnected as a porous monolithic support which can improve charge carrier lifetime compared to assemblies or powders.

In this work, we developed a metal-semiconductor photocatalyst that uses a synergy of LSPR at the Au NP surface, electronic absorption of the semiconductor support, and slow photon effects near a pseudo photonic band gap of an inverse opal (IO) photonic crystal[30] to enhance light absorption (white light or monochromatic) for plasmonically enhanced photocatalysis. Plasmonic photocatalysts are prepared by depositing monodisperse Au NPs on a photonic crystal $V_2O_5$ support with an IO structure. $V_2O_5$ is a visible light responsive, photocatalyst with an optical bandgap of $E_g$ ~2.3 eV. $V_2O_5$-graphene nanostructures showed strong degradation efficiency of dyes with direct sunlight irradiation[31] and coupling $V_2O_5$ with $TiO_2$ or $SnO_2$ has been shown to improve photocatalytic efficiency.[32] Catalytic performance is compared with Au-$TiO_2$ catalysts by depositing Au NPs onto $TiO_2$ IOs with an bandgap of ~3.2 eV (for anatase form). The catalytic activity for hydrogenation of nitrophenol is investigated under different illumination conditions using broadband UV-visible light, monochromatic green light ($\lambda$ = 532 nm) and under no illumination. The highest catalytic enhancement was achieved when Au NPs were deposited on a visible light responsive semiconductor, $V_2O_5$ IO catalyst under green light excitation due to spectral overlap of the electronic band gap, LSPR and excitation source. We further demonstrate a significant influence of the semiconductor superstructure by comparing non-porous supports with photonic crystal supports which enable photonic band gap (PBG) and slow light effects for further photocatalytic enhancement resulting in a doubling of the reaction rates for both $V_2O_5$ and $TiO_2$ IO structures compared to non-porous catalysts.

**Experimental**

All chemicals used unless otherwise stated were purchased from Sigma-Aldrich and used as received.

*Au nanoparticle synthesis:* Au NPs were prepared using a previously published method.[33] Briefly, $HAuCl_4$ (200 mg) were combined with OAm (20 mL) and 20 mL of 1,2,3,4-tetrahydronaphthalene (20 mL) in a round bottom flask. *t*-Butylamine-borane complex (86 mg, 1 mmol) dissolved in of OAm (2 mL) and of 1,2,3,4-tetrahydronaphthalene (2 mL) was quickly injected into the solution. The solution was stirred in air for 1 h. The particles were precipitated using ethanol and centrifugation at 8000 rpm for 20 min for three cycles before being redispersed in 20 mL of toluene.

*Inverse opal synthesis and NP immobilization*: Synthesis of $V_2O_5$ and $TiO_2$ IO on FTO was carried out using a previously described procedure[34, 35]. Briefly, opal templates were formed by electrophoretically depositing ~500 nm PS spheres on FTO-glass substrates, cleaned with acetone, IPA and deionized water. A 50:1 ratio of IPA to vandadium triiospropoxide oxide ($OV(OCH(CH_3)_2)_3$) was added to a 500:1 IPA-deionized water mixture and stirred until clear, forming a 1000:10:1 precursor solution. The precursor was then drop casted onto opal templates and annealed in an oven at 300°C for 12h to remove the template. $TiO_2$ IO's were synthesized by dissolving $TiCl_4.2THF$ (334 mg) in IPA, forming a 0.1M precursor solution. Samples were annealed at 450°C for 1h. $V_2O_5$ and $TiO_2$ non-porous thin films were formed by drop-casting the IPA-precursor solutions into cleaned FTO substrates to obtain a uniform thin film coating. $V_2O_5$ and $TiO_2$ samples were crystallized by annealing at 300°C (12 h) and 450°C (1h), respectively. Au nanoparticle immobilization was carried out by immersing the substrates in Au NP solution (1 mg ml$^{-1}$) overnight. The substrates were removed and left to dry. Substrates were rinsed with toluene to remove excess Au NPs and dried in air.

*Reaction studies:* Catalytic reduction of 4-nitrophenol (TCI Chemicals) was investigated using in-situ UV-Vis spectroscopy. The UV-vis spectral analysis was carried out using a quartz tungsten-halogen lamp operating from 400 - 2200 nm from Thorlabs Inc., a UV-Visible spectrometer (USB2000+ VIS-NIR-ES) with operational range 350 – 1000 nm from Ocean Optics Inc. A motorised rotation stage (ELL8; Thor Labs Inc.) was used to control the incidence angle in transmission measurements. Laser excitation at $\lambda$ = 532 nm was supplied using a Laser Quantum GEM DPSS single transverse mode CW laser and focused using an objective lens. For catalysis studies, the reaction rate was determined by monitoring the decrease in absorption of 4-NP at $\lambda$ = 400 nm. The apparent rate constant $k_{app}$ was estimated from the slope of $-\ln(A/A_0)$ vs time. In a typical reaction, 2 ml of deionized water, 0.25 ml of a 0.5 mM of 4-nitrophenol solution and a stir bar were added to a polystyrene cuvette. The reaction was initiated by addition of 0.5 ml of freshly prepared 0.2 M $NaBH_4$ solution.

*Materials characterization:* Scanning Electron Microscopy (SEM) characterization was performed using a Hitachi S-4800 SEM cold field emission apparatus or a SU-70 SEM hot field emission apparatus. Energy dispersive X-ray (EDX) spectroscopy was carried out using an Oxford X-Max 80 detector. Transmission Electron Microscopy (TEM) was carried out using a JEM2010-TEM. X-ray Photoelectron Spectroscopy (XPS) was acquired using a KRATOS AXIS 165 monochromatized X-ray photoelectron spectrometer equipped with an Al K$\alpha$ ($h\upsilon$ = 1486.6 eV) X-ray source. Spectra were collected at a take-off angle of 90° and all spectra were reference to the C 1s peak at 284.6 eV. The spectra were fit to Voigt profiles using a Shirley background.

*Electrical characterization:* Electrical conductivity measurements were performed using a Keithley Instruments 2612B Dual-Channel System Sourcemeter, with gold-coated probes and In-Ga eutectic as a contact. Light sources placed across contacts were a $\lambda$ = 532 nm excitation source operating at 20 mW and a tungsten halogen lamp operating at 0.8 mW. Electrical transport measurements of $TiO_2$ were performed between -10 V and +10 V. Measurements of $V_2O_5$ materials were performed between -4 V and +4 V, with 500 data points, and averaged over 3 repeatable I-V curves for each sample.

*Modelling of photonic bandgap and slow photon effects:*

Finite different time domain (FDTD) models were used to compute the photonic band structure of $TiO_2$ an $V_2O_5$ inverse opal photonic crystals in the first Brillouin zone. Physical dimensions were acquired for statistical analyses of IO pore diameters and periodicity, and the full details and data are provided in Supporting Information. To gauge slow photon effects in IOs, the group indices $n_\mathrm{g} = c/\upsilon_\mathrm{g}$, where $\upsilon_\mathrm{g} = \frac{\partial \omega}{\partial k}$, and optical path length $L = n_\mathrm{eff} \cdot s$ at $\Gamma - \mathrm{M}$ and $\mathrm{M} - \mathrm{K}$ band edges were calculated, where $n_\mathrm{eff}$ is the effective index of the solution-filled IO material in each case for both TE and TM polarizations.

**Results and discussion**

*Plasmonic photonic crystal photocatalysts*

**Figure 1** summarizes four characteristics we exploit to maximize photocatalytic activity: Hot electron effects at the surface of Au NPs under white and green light illumination, the photonic band gap of the oxides to maximize absorption, slower group velocity for photons in the energy range of interest, and control over electron transfer from Schottky barriers between the Au NP and each oxide. Plasmonic photonic crystal photocatalysts were prepared using a stepwise approach by combining inverse opal (IO) thin film coatings with monodisperse colloidal Au NPs. Firstly, IO thin films of $V_2O_5$ and $TiO_2$ were prepared on ITO substrates by infiltration of liquid precursors into polystyrene templates (see Experimental section for details). The resulting IO photonic crystal structures were surface functionalized with Au NPs. **Figures 2 (a) and (b)** show SEM and TEM analysis of the $V_2O_5$ IO, with the internal walls of the IO framework consisting of a layered structure, typical of orthorhombic $V_2O_5$ which is a vdW layered material. Further details can be seen in Supporting Information **Figures S1 and S2**. XRD of the $V_2O_5$ and $TiO_2$ IOs (**Figure S3**) confirms crystalline orthorhombic $V_2O_5$ and anatase $TiO_2$ of the materials in IO form.[35] In comparison, the morphology of $TiO_2$

IOs is comprised of interconnected TiO$_2$ NPs, as shown in **Figure 2 (c) and (d)** with internal pore dimeters that are ~460 nm in diameter.[36] Similar pore dimensions that are also found on the V$_2$O$_5$ IO structures (**Figure S4**). A colloidal solution of monodisperse Au NPs with a mean diameter of 4.5 nm, shown in **Figure 2 (e)**, were immobilized onto the IO supports giving well-dispersed Au NPs across the IO as shown in **Figure 2 (f)**. Using Au NPs dispersed in low dielectric constant solvent such as hexane, coupled with a large PS template (500 nm) allows the NP solution to percolate through the IO support.[37] SEM image and EDX analysis shown in **Figure 2 (g) and (h)** and in **Figure S5**, confirms Au decoration of the V$_2$O$_5$ and TiO$_2$ IO supports. Non-porous Au-NP decorated thin films of each oxide were also prepared to determine the influence of the IO structure on catalytic performance.

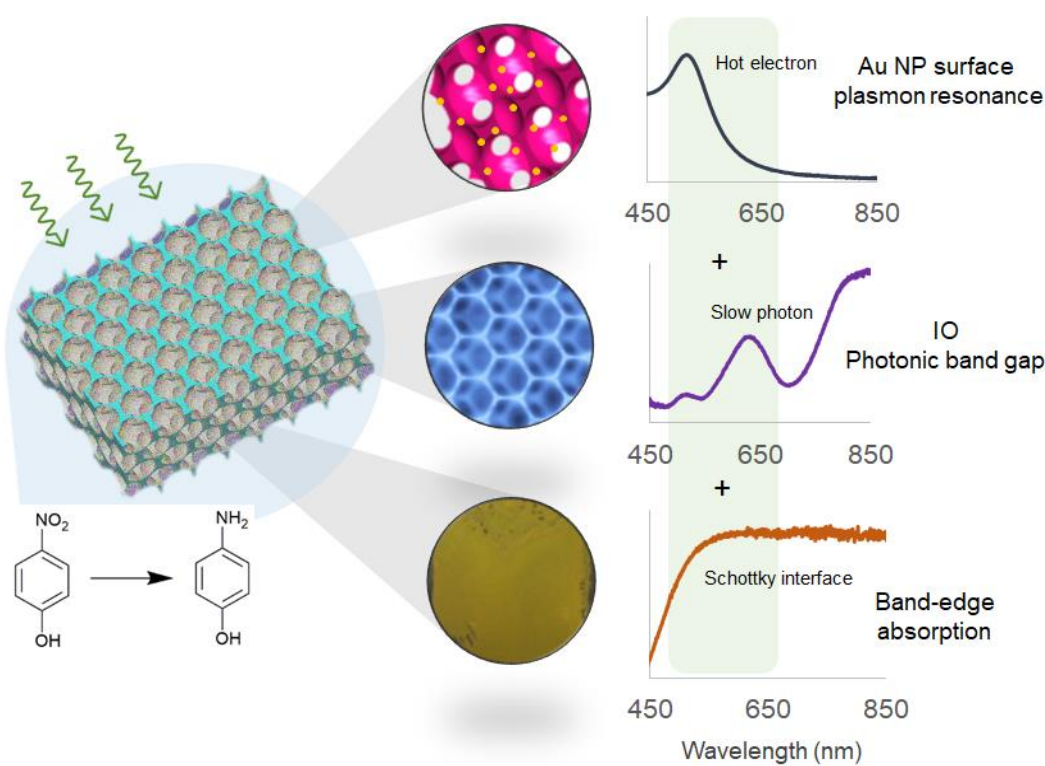

**Figure 1.** Representation of semiconductor photonic crystal plasmonic photocatalyst design and synergy. Inverse opal photonic crystals functionalized with Au NPs photocatalytically reduce nitrophenol under green (532 nm) laser light or broadband white light. Both TiO$_2$ and V$_2$O$_5$ are examined, with band gaps in the UV (~3.2 eV) and visible regions (~2.3 eV) respectively. The system involves surface plasmon-mediated hot electrons at the Au surface, band-edge absorption from the semiconducting metal oxide, Schottky barrier from semiconductor-metal interface, and pseudo-photonic band gap (including slow photon effects) from the inverse opal, to tune light absorption and electron transfer during photocatalysis.

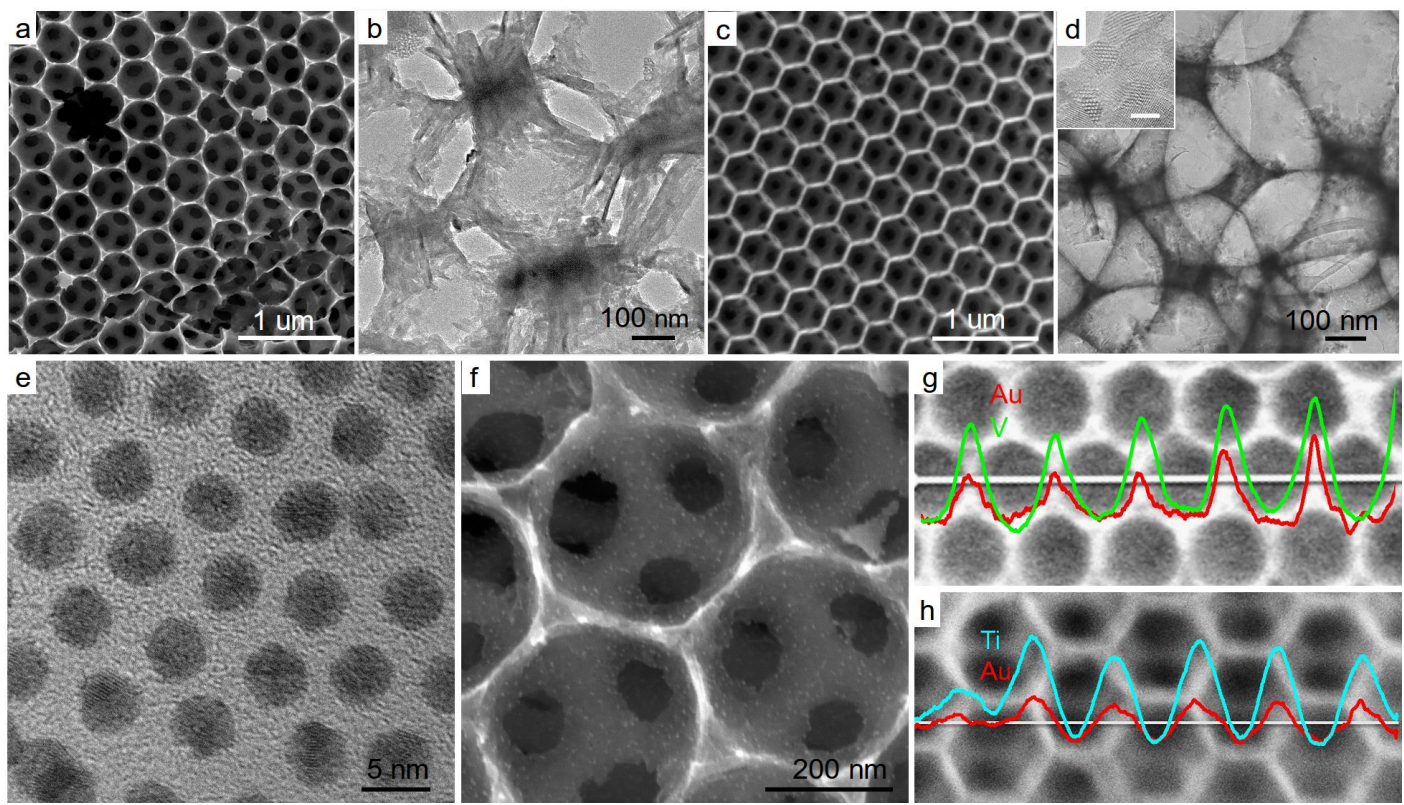

**Figure 2.** (a) SEM and (b) TEM images of $V_2O_5$ inverse opals and the vdW layered internal structure of the vanadate IO walls, and the infilling of the tetrahedral and octahedral voids of the parent opal template. (c) SEM and (d) TEM images of the $TiO_2$ IO material. The inset shows HRTEM images of the Au NP-$TiO_2$ IO interface after NP functionalization. (e) TEM image of synthesized Au NPs. (f) $V_2O_5$ IO after Au NP immobilization. (g) EDX line scan of $V_2O_5$ and (h) $TiO_2$ IO photocatalysts decorated with immobilized Au NPs.

**Figure 3(a) and (b)** show the V 2p and Ti 2p core level spectra of the IO supports, respectively. The V 2p spectrum confirms the presence of $V^{5+}$ oxidation state with peaks for the V 2p doublet at binding energies (B. E.) of 517.7 eV and 525.2 eV, assigned to $V^{5+}$ $2p_{3/2}$ and $V^{5+}$ $2p_{1/2}$, respectively. The O 1s signal in the same spectrum shows photoemission at 530 and 532.2 eV, attributed to the surface lattice oxygen and adsorbed oxygen species, respectively.[38] The Ti 2p spectrum shows the doublet at 464.8 and 459.2 eV assigned to Ti $2p_{1/2}$ and Ti $2p_{3/2}$, respectively. From the Au 4f spectra is shown in **Figure 3(c)** the B. E. of the unsupported Au NPs at 83.6 eV is positively shifted to 84.4 eV when immobilized on $V_2O_5$ and negatively shifted to a lower binding energy of 83.4 eV when immobilized on $TiO_2$. A B. E. difference of +1.2 eV between the two support materials is indicative of Au being more electron deficient when deposited on $V_2O_5$ compared to $TiO_2$. This charge transfer can be rationalized by comparison of the relative work functions (WF) of the two semiconductor supports.[39]

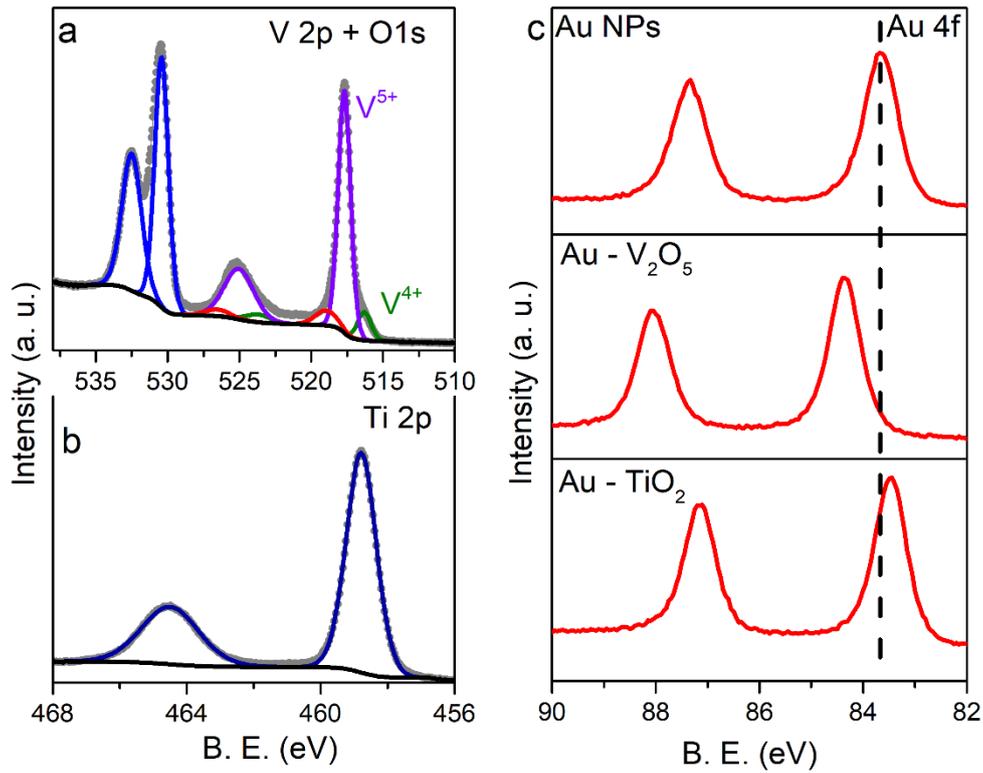

**Figure 3.** (a) XPS core level spectra of V 2p and O 1s and (b) Ti 2p. (c) Comparison of Au 4f core-level spectra of Au NPs immobilized on $V_2O_5$ and $TiO_2$ IO.

To tune the synergy of material and optical properties, we first characterized the spectral overlap of the LSPR for Au NPs and the photonic bandgap and band-edge absorption for the $V_2O_5$ and $TiO_2$ IOs. The optical band gap energies, IO photonic band gap ranges and absorption resonance of the Au NPs were determined for the $V_2O_5$ and $TiO_2$ IOs with and without Au NP functionalization. **Figure 4(a) and (b)** show the absorption spectra for Au NPs in solution, confirming an LSPR of ~530 nm. Transmission spectra of the IOs at normal incidence show the location of the band-edge, and the stop band associated with a pseudo photonic band gap, effectively extending the energy window for photon absorption for white light. Optical bandgaps were estimated from UV-vis absorption spectra and analysed in the framework of the Tauc model. **Figure 4(c,d)** show the plots of $(\alpha h\upsilon)^2$ versus $h\upsilon$ where $\alpha$ is the absorption coefficient near the absorption edge, $h$ is Planck's constant and $\upsilon$ is the photon frequency. The extrapolated band gap values were 2.3 eV for $V_2O_5$ and 3.2 eV for $TiO_2$, in good agreement with literature values.[40] In the case of $TiO_2$ with a band gap in the UV, the Au decorated $TiO_2$ IO absorption profile changes to show an sub-bandgap Urbach, further indicating light absorption extending into the visible region associated with the surface plasmon absorption from the Au NPs. Visible light adsorption is also further enhanced by the presence of the a PBG associated with the IO structure (**Figure 4(a)**). The valence band position was evaluated from photoelectron valance band (VB) spectra of the catalysts before and after Au NP deposition. The VB density of states of bare $V_2O_5$ (**Figure 4 (e)**) shows the band edge located at 2.9 eV and after Au NP deposition the main absorption on-set in the VB spectrum shifts to 2.0 eV. The VB density of states of bare $TiO_2$ (**Figure 4 (f)**) shows the band edge is 2.7 eV and after Au deposition spectrum is blue shifted to ~2 eV.

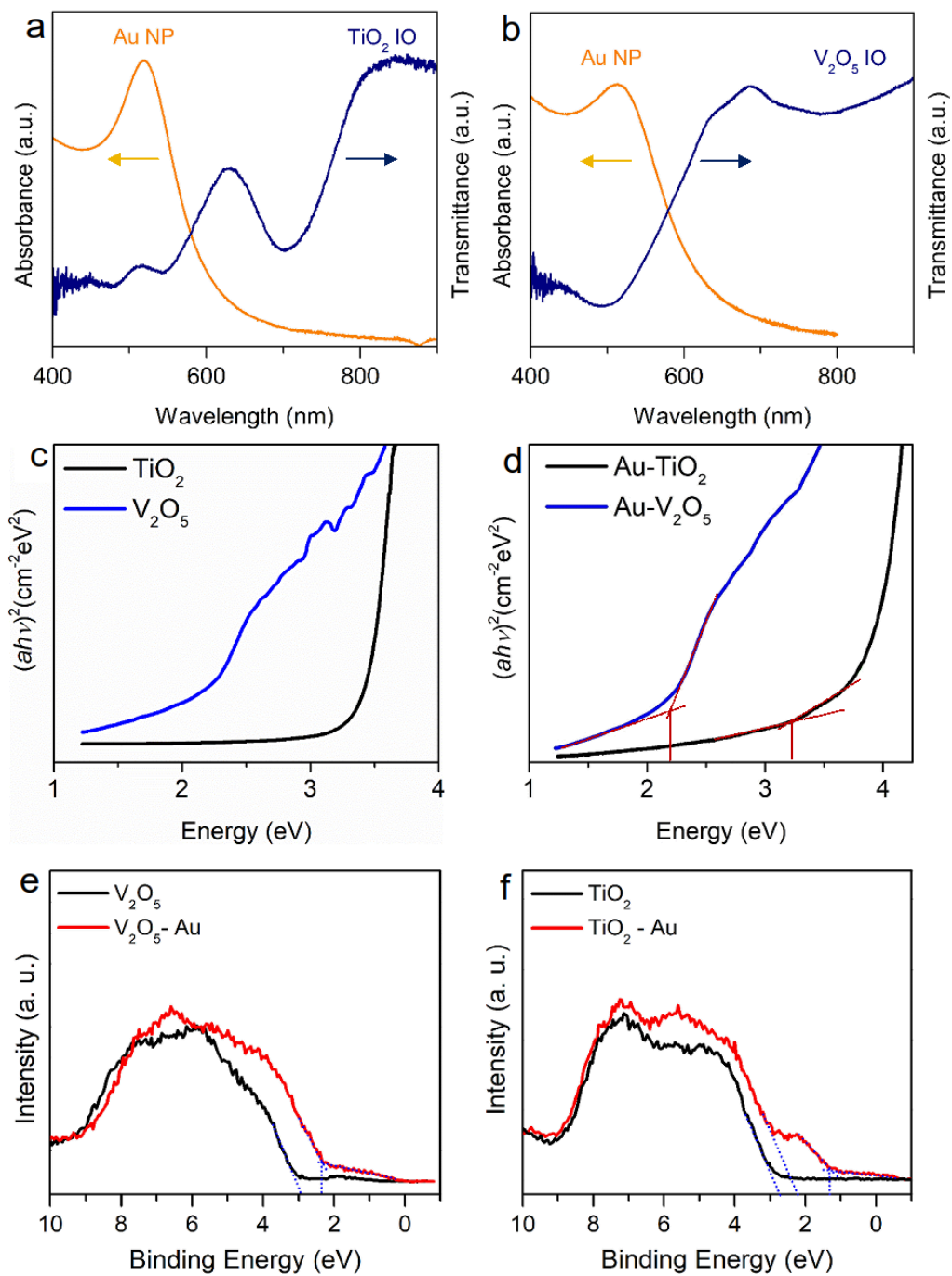

**Figure 4.** Spectral absorbance and LSPR of (a) Au NP in hexane and the transmittance of the $TiO_2$ IO acquired at normal incidence, and (b) Au NP LSPR with transmittance of the $V_2O_5$ IO. In both IOs, the pseudo PBG (stop band) can be clearly observed at lower wavelengths. The relative contribution of the stop band in $TiO_2$ to electronic absorption is more significant; $V_2O_5$ IO transmittance is dominated by electronic absorption, with a broad band reflectance from the PBG of the IO structure. Tauc plots of (c) bare and Au-decorated $V_2O_5$ IO catalyst and (d) bare and Au-decorated $TiO_2$ IO catalysts converted from UV-vis transmission data. (e) Valance band (VB) spectra from XPS of bare and Au decorated $V_2O_5$ IO catalyst and (f) corresponding VB spectra from bare and Au decorated $TiO_2$ IO catalysts.

*Photocatalytic enhancement from synergy of LSPR, PBG and band-gap absorption*

Electron transfer of plasmonic NPs coupled with high bandgap semiconductors, such as Au-TiO$_2$ systems, is generally attributed to plasmonic sensitization.[41] A Schottky barrier, $\varphi_B = \varphi_M - \chi$, is formed at the interface where $\varphi_M$ is the workfunction of Au and $\chi$ is the oxide semiconductor electron affinity. Energies for hot electrons generated from excitation of the LSPR typically range from 1-4 eV, can then transfer into the TiO$_2$ CB, or the lowest unoccupied molecular orbital LUMOs of molecular adsorbates[42] as illustrated in **Figure 5**. The Schottky barrier height estimated using values of $\varphi_M$ = 5.1 eV for Au, $\chi_{TiO_2}$ = 4.0 eV and $\chi_{V_2O_5}$ = 6.3 eV,[43] correspond to, 1.1 eV and -1.2 eV, respectively, thus favouring electron transfer to the Au NP from the V$_2$O$_5$ CB and vice versa for TiO$_2$. In the case of V$_2$O$_5$, which is a visible light responsive semiconductor, light simultaneously excites the LSPR at the NP and electron/hole pair generation in the V$_2$O$_5$. The energy level of the V$_2$O$_5$ CB is higher than the Fermi level of the Au NPs and so excited CB electrons generated in V$_2$O$_5$ should favor transfer to the Au NPs, as shown in **Figure 5**. Furthermore, the LSPR is resonant with the V$_2$O$_5$ band edge which should further contribute to enhancing photocatalytic performance.

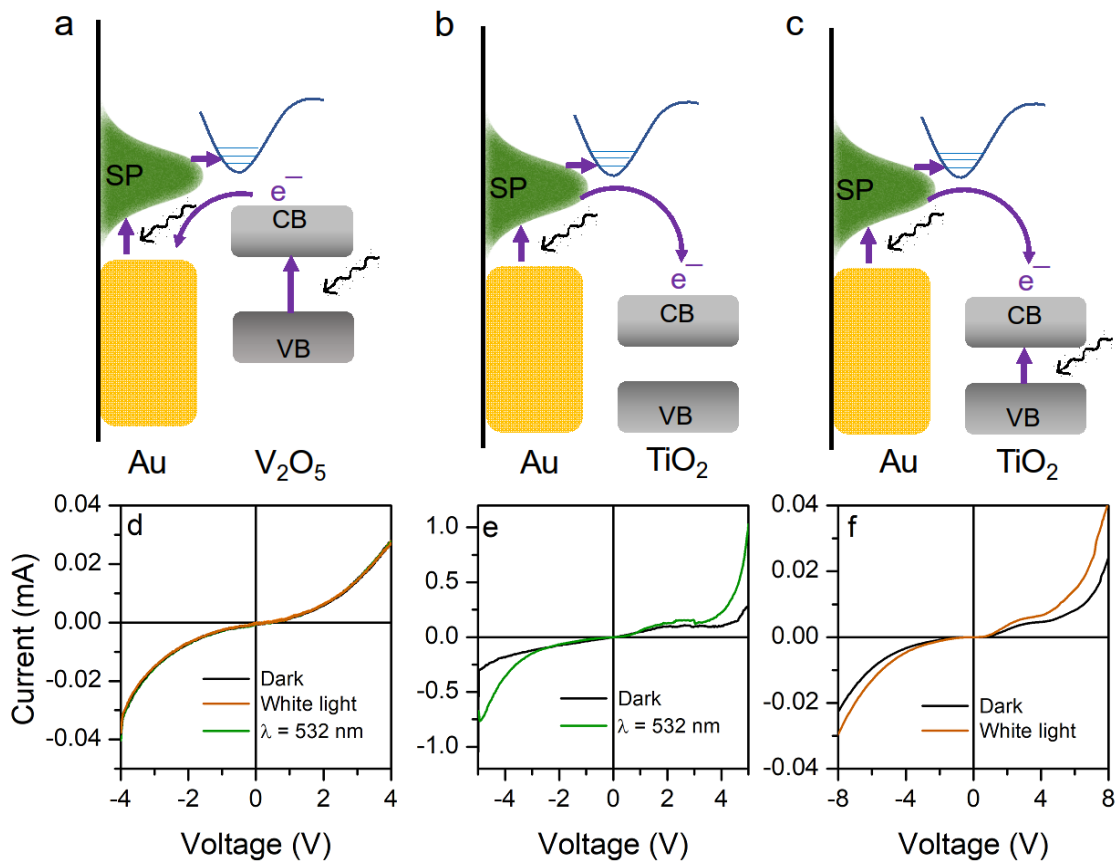

**Figure 5.** Band structure comparing Au NP supported on (a) V$_2$O$_5$ under white light and $\lambda$ = 532 nm (b) TiO$_2$ under $\lambda$ = 532 nm irradiation and (c) TiO$_2$ under white light illumination. SP represents surface plasmons, CB is conduction band, and VB is valence band. I-V curves acquired at room temperature of Au NP-functionalised (d) V$_2$O$_5$ and (e) TiO$_2$ IO materials under halogen white light and (f) a separate Au-TiO$_2$ IO excited by $\lambda$ = 532 nm photons.

To further examine the nature of the photocurrent generation and electron transfer in Au NP-functionalized IOs, electrical transport measurements (in the dark and under illumination) were conducted. **Figure 5 (d-f)** shows two terminal I-V curves for Au-$V_2O_5$ and Au-$TiO_2$ IOs under white light, excitation at $\lambda =$ 532 nm, and in the dark. The Au-NP functionalized IO structures display typical behaviour consistent with Schottky barriers to the n-type oxide from the In-Ga eutectic contacts. The Au-$V_2O_5$ IO showed negligible photocurrent response using either light source as shown in **Figure 5(d)**. With a bandgap of 2.3 eV, $V_2O_5$ would be expected to exhibit some photoresponse from both green and white light illumination in the form of a weak Schottky photodiode. This observation is attributed to extraction of electrons to an outer circuit to register as photocurrent is limited by direct and efficient electron transfer from the $V_2O_5$ CB to the Au NPs in this case. By contrast, the photocurrents of the Au-$TiO_2$ IO improved upon illumination with a halogen UV-visible light source and under $\lambda =$ 532 nm excitation, as shown in **Figure 5(e) and (f)**. In these cases, green light excites the Au NP SPR which then, as illustrated in **Figure 5(b,c)**, transfer to the $TiO_2$ CB, giving the observed photoresponse. Increased photocurrent is also observed under broad band visible light attributed both excitation of the SPR and to VB-CB transitions in $TiO_2$ from the UV-portion of the white light source exciting the band gap (3.2 eV). Efficient electronic coupling and fast electron transfer between Au NPs and $TiO_2$ supports has been confirmed by femtosecond transient absorption spectroscopy with an infrared probe.[44]

*Quantifying the enhancement from photonic crystal plasmonic photocatalysts*
Testing the synergy between LSPR, band-edge electronic absorption, and maximizing optical path length and absorption using a plasmonic-photonic crystal architecture was done by exploring catalytic reduction of 4-nitrophenol (4-NP) by $NaBH_4$, which is often used as a model reaction to evaluate the behaviour of metal NPs.[45] The reaction proceeds via the intermediate 4-hydroxylaminophenol, and requires that both reactants (4-NP and $BH_4^-$) must first absorb on the metal surface. The apparent rate constant $k_{app}$ was estimated for the Au-decorated $TiO_2$ and $V_2O_5$ IO catalysts from in situ UV-vis spectroscopy (**Figure S6**). **Figure 6** compares the catalytic and photocatalytic activity of the IO substrates under illumination with (i) a broad band white light halogen lamp, (ii) green laser excitation at $\lambda =$ 532 nm and (iii) in the dark (see Experimental Section for details). The $k_{app}$ for Au-$V_2O_5$ IO without any illumination was 1.26 × $10^{-3}$ $s^{-1}$, and this rate almost doubled to 2.2 × $10^{-3}$ $s^{-1}$ under visible light irradiation. On excitation with green light at $\lambda =$ 532 nm, the rate increased by an order of magnitude with $k_{app} = 1.16 \times 10^{-2}$ s$^{-1}$. The large $k_{app}$ when the excitation wavelength is coincident with the LSPR is indicative of plasmonic photocatalysis being responsible for higher reaction rate under illumination. **Figure 6(b)** shows the catalytic performance of the Au-TiO$_2$ IO catalyst, which also displayed photocatalytic enhancement but to a lesser degree compared to Au-V$_2$O$_5$. Under no illumination, the $k_{app}$ estimated for the Au-TiO$_2$ IO was $8.25 \times 10^{-4}$ s$^{-1}$, ~60% lower than that of the V$_2$O$_5$ catalyst, demonstrating the Au-V$_2$O$_5$ system to be a general superior catalyst for this reaction. A longer induction period of ~60 s was observed for the reactions catalyzed by the Au-TiO$_2$ IOs. Induction periods are sometimes observed for this reaction and have been associated with surface restructuring of the NP before the adsorption of the reagents.[46] As the same colloidal Au NPs were used for both TiO$_2$ and V$_2$O$_5$, charge transfer at the metal-semiconductor interface, as identified by the shifts in the Au 4f core-level B.E., may contribute to the longer induction period observed for TiO$_2$. Photocatalytic enhancement for the Au-TiO$_2$ IO was highest under UV-visible light irradiation, with the reaction rate increasing by ~70% ($k_{app} = 1.37 \times 10^{-3}$ s$^{-1}$). Interestingly, despite the green light excitation being coincident with the Au SPR, the photocatalytic activity of the Au-TiO$_2$ IO under $\lambda = 532$ nm ($k_{app} = 1.19 \times 10^{-3}$ s$^{-1}$) was lower than under halogen light irradiation. As shown in the absorbance spectra (**Figure 3(a)**) while the LSPR is not resonant with the TiO$_2$ band edge, the Au-TiO$_2$ IO catalyst does show strong visible light adsorption associated with the PBG, thereby leading to improved catalytic performance under visible light. The UV component of the halogen light source which can excite the TiO$_2$ band gap may also play a role in the enhanced activity.

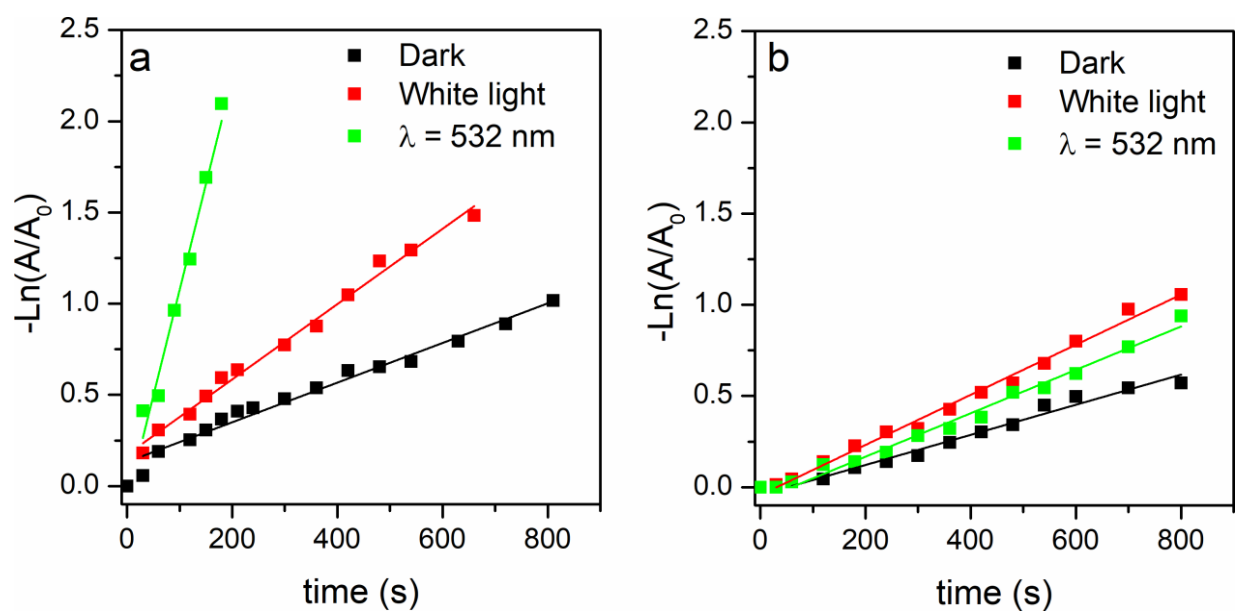

**Figure 6.** Reaction profile for 4-nitrophenol reduction from Au NP functionalised IOs of (a) V$_2$O$_5$ and (b) TiO$_2$ under white light illumination, $\lambda = 532$ nm excitation, and in the dark.

To further study the slow photon effect in these photonic crystal photocatalyst systems, the nature of slow group velocity photons in the $TiO_2$ and $V_2O_5$ IOs immersed in the solution ($n_{sol} = 1.45$) was evaluated. The IOs are designed such that the optical path length increase from slow photons at the higher energy edge of the photonic band gap, where the electric field is localised within the higher index material (metal oxide).[47] Using the optical transmission data acquired at normal incidence (see **Figure S7**), we calculated the respective group index $n_g = c/v_g = \frac{d\omega}{dk}$ for both IOs, shown in **Figure 7**. More information on the finite difference time domain models can be found in Supporting Information and in **Figures S8 and S9**. Using a Drude approximation, $n_{eff}$ = 1.82 and 1.56 for $TiO_2$ and $V_2O_5$ IOs, leading to reduced group indices shown in **Figure 7(a) and (b),** which illustrates that photon absorption across the visible range is enhanced by the slow photon effect in both $TiO_2$ and $V_2O_5$ IOs catalysts. For $V_2O_5$, the optical path length $L = n_{eff} \cdot s$ where $s$ is the geometrical length, is a factor of 1.2× greater for $V_2O_5$ IO at $\lambda$ = 532 nm.

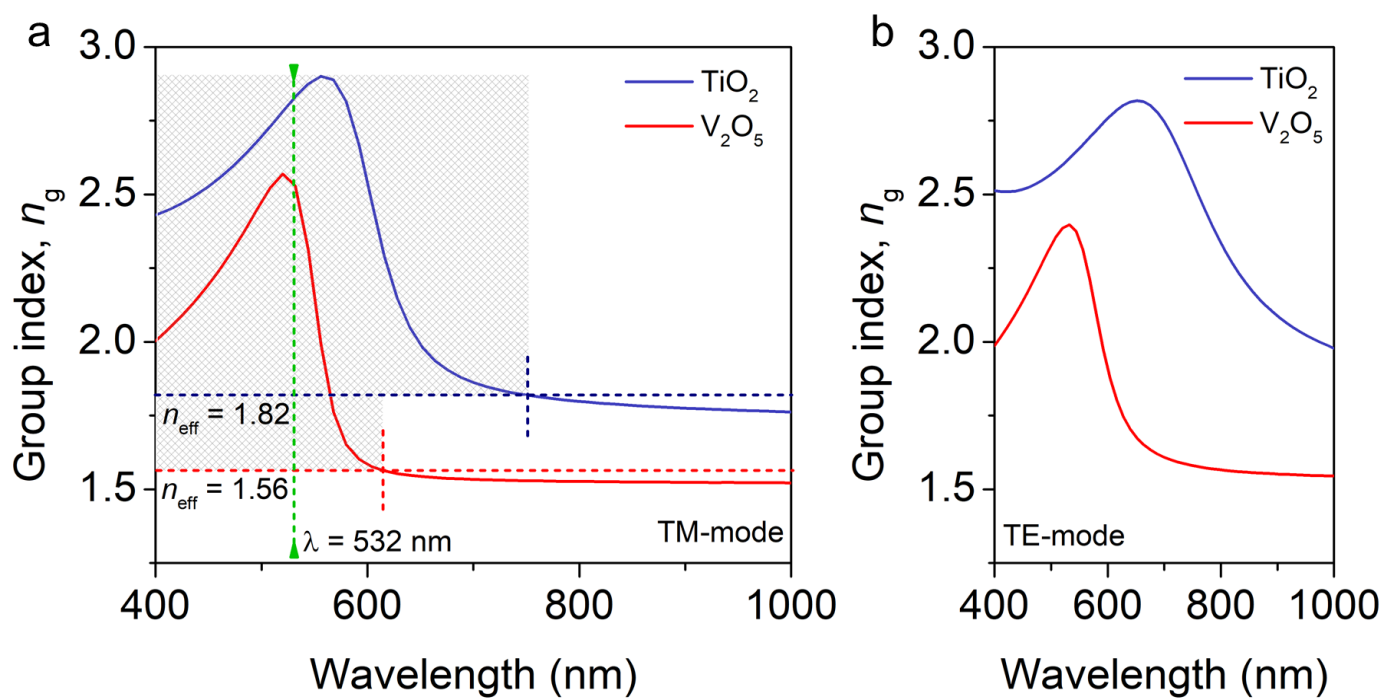

**Figure 7.** Computed variation of the group index $n_g$ for reaction solution-infiltrated $TiO_2$ and $V_2O_5$ IO photonic crystal photocatalysts in (a) TM and (b) TE polarizations. The regions for higher group index (slow photon group velocity $n_g = c/v_g > n_{eff}$) are shaded in (a). The effective index $n_{eff}$ for the solution-filled IO medium for each case are shown by the dashed horizontal lines.

Notably, over the entire spectralrange examined in **Figure 7**, $TiO_2$ has a longer effective optical path length (higher group index from slow light) at all frequencies under TM or TE polarizations. This is relevant for enhanced photon absorption and photocatalysis under white light illumination, which correlates well to the superior activity observed under broadband visible light. For example, the maximum effective increase in group index for TiO$_2$ (**Figure 7(a)**) at $\lambda$ = 558 nm is a factor 1.52× greater than for the V$_2$O$_5$ IO. The maximum enhancement effect (~1.75×) is found at $\lambda$ = 580 nm.

The behaviour of the metal-semiconductor hetero-structured catalysts under broad-band white light is of particular interest for harvesting solar energy for sustainable catalysis and so the photocatalytic rate behaviour was evaluated further. To study the inherent structural features of the IO architecture, and synergy between the oxide bandgap, PBG and slow light effect as demonstrated in **Figure 7**, non-porous catalysts of TiO$_2$ and V$_2$O$_5$ were prepared using the same procedure as the IOs but in the absence of the IO structure-defining polystyrene template. XRD patterns (**Figure S1**) and AFM images of the non-porous catalytic thin films are shown in **Figure S10**, confirming the same stoichiometric crystalline phase of V$_2$O$_5$ and TiO$_2$ in both cases. **Figure 8 (a) and (b)** compare the reaction rate profiles of the Au-V$_2$O$_5$ and Au-TiO$_2$ IO supports, and Au NP-decorated non-porous supports, respectively. Band alignment, LSPR and band-edge absorption are nominally similar in both sets of catalyst systems, with exception of the IO structure. Under visible light the V$_2$O$_5$ catalyst with an IO structure displayed a doubling of the rate compared to the non-porous V$_2$O$_5$ catalysts. Similar enhancements were observed for the Au-TiO$_2$ IO catalysts compared to the non-porous TiO$_2$ catalyst, indicating the support architecture plays a key role in the photocatalytic enhancement. This rate enhancement can be attributed to the slow photon effect and the ordered IO superstructures which increase charge separation, prolonging the lifetime of charge chargers.[10] The reaction data in **Figure 8 (a) and (b)** further demonstrates that both V$_2$O$_5$ and TiO$_2$ catalysts having an IO architecture display enhanced performance even under no illumination, which originates from the ordered macroporous IO structure which is known to be beneficial for liquid and vapour phase reactions due to efficient mass transport and wetting of surfaces to allow infiltration of reaction species.[48]

The bar chart in **Figure 8(c) and (d)** compares reaction performance of catalysts having an IO and non-porous structure under different illumination conditions (broadband vs monochromatic). Within these Au-semiconductor catalysts the catalytic enhancement can originate from a synergy of effects (*cf*. **Figure 1**) associated with (i) the LSPR of Au NPs at 520 nm, (ii) Schottky barrier-mediated charge-transfer at Au–semiconductor interface, and (iii) the role of the IO superstructure and PBG compared to non-porous supports. Catalytic enhancement associated with coupling these effects is clearly demonstrated for the Au-V$_2$O$_5$ IO under excitation at $\lambda$ = 532 nm, when the V$_2$O$_5$ electronic band gap (2.3 eV) overlaps with the Au NP LSPR (~530 nm) in the same energy region as the lower photon group velocity of the photonic band gap (slow light effects), resulting in an order of magnitude rate increase. The rate enhancement obtainable in the Au-$V_2O_5$ catalyst under UV-vis light was lower by a factor of five without these synergistic effects. Specifically engineering the energy range to use LSPR, oxide bandgap and PBG (including light trapping) using slow photon effects, significantly enhances plasmonic photocatalysis from Au-$V_2O_5$ IO where charge transfer is optimized to the catalysing Au NP interface.

In the case of the Au-$TiO_2$ catalyst, plasmonic enhancement is mainly limited to plasmonic sensitization due to the wider band gap of $TiO_2$ (3.2 eV). The relative work-functions and associated band bending promote electron injection into the $TiO_2$ CB, the opposite charge transfer observed for the Au-$V_2O_5$ catalyst, as illustrated in **Figure 5**. For the Au-$TiO_2$ catalyst despite the excitation wavelength at $\lambda$ = 532 nm being nea-resonant with the LSPR, a greater photocatalytic enhancement is observed under broadband halogen light, attributed with the presence of the PBG enabling visible light adsorption (**Figure 3**) and associated slow photon effects (**Figure 7**). Furthermore, the UV component of the halogen light can generate electron-hole pairs in addition to hot electrons from the Au into the CB which may also contribute to the enhanced photocatalytic activity observed. Although the exact mechanism of this catalytic behaviour is unclear, several studies demonstrate the multi-faceted nature with positive and negative effects arising from the LSPR in Au/$TiO_2$ structures under different illumination sources. [49]

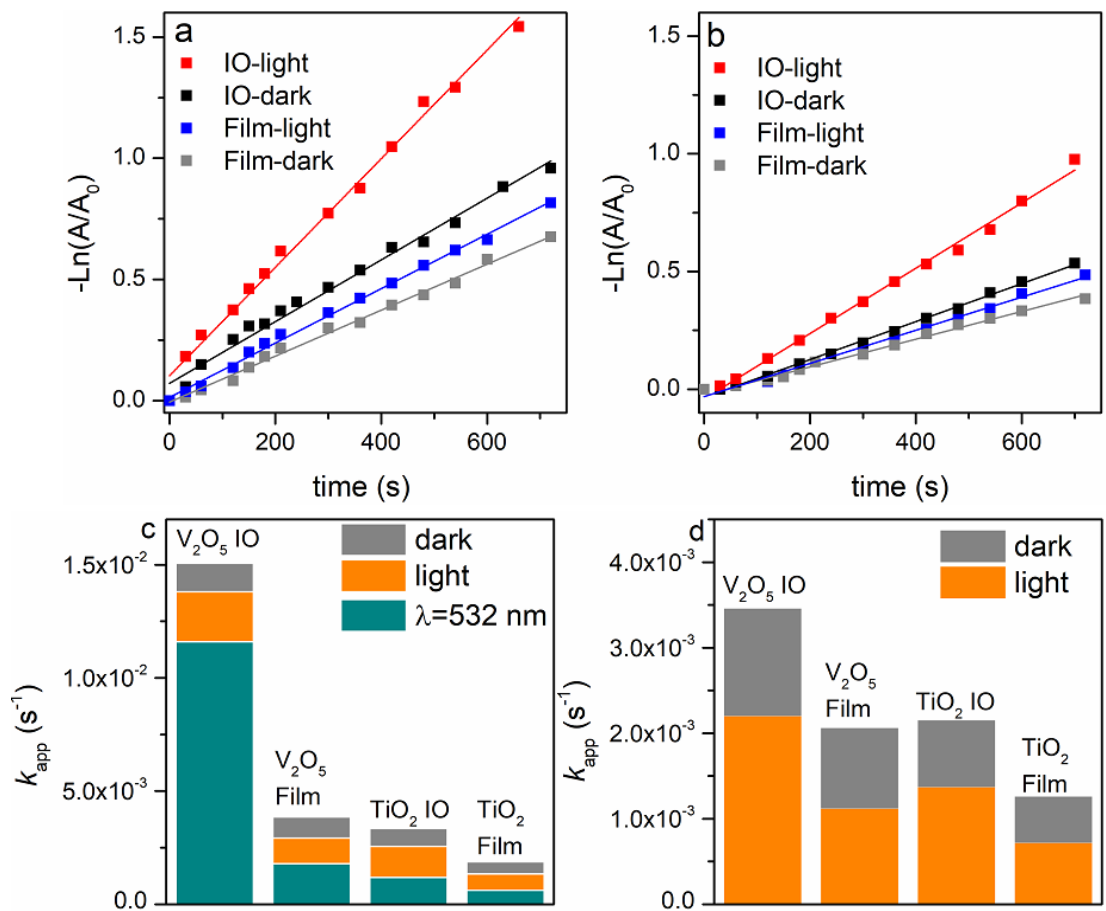

**Figure 8.** Reaction profiles for plasmonic photocatalytic 4-nitrophenol reduction using IO and non-porous thin films (a) $V_2O_5$ and (b) $TiO_2$ catalysts in the dark and under visible light and monochromatic (green) irradiation.

**Conclusions**

Plasmon-enhanced photocatalysis holds significant promise for enhanced performance and controlling chemical reaction rates. Semiconductor photocatalysts with an inverse opal structure offers a large active surface area, electrically interconnected porous network which can be functionalised with plasmonic NPs to form plasmonic-photonic catalysts with broad visible absorption due to the presence of a PBG to further enhance the efficiency. The use of $V_2O_5$ as a visible light semiconductor catalyst allowed several synergetic effects including the LSPR, electronic bandgap, PBG and slow photon effects, resulting in superior performance to a conventional $TiO_2$ support for hydrogenation of 4-nitrophenol. Both $V_2O_5$ and $TiO_2$ IO catalysts have superior photocatalytic compared to non-porous catalysts due to integration of the Au SPR and the PBG associated with the inverse opal structure of the support. Overall, this strategy takes many of the chemical, material and photonic strategies used to control photon-to-electron conversion for photocatalysis in a synergistic way to improve visible light operation. The modularity of the synthesis approach facilitates rational design of efficient plasmonic photocatalysts as the nanoparticle and semiconductor components can be readily altered, enabling it to be extended to other metal-semiconductor composites for a variety of catalytic applications.

**Supporting Information**

Supporting Information is available from the Wiley Online Library or from the authors.

**Acknowledgments**

AL acknowledges support from the Irish Research Council Government of Ireland Postgraduate Scholarship under award no. GOIPG/2016/946. DB acknowledges support from the Irish Research Council Government of Ireland Postgraduate Scholarship under award no. GOIPG/2014/206. COD acknowledges funding support from Science Foundation Ireland (SFI) under Awards no. 14/IA/2581 and 15/TIDA/2893, and from the Irish Research Council Advanced Laureate Award under grant no. IRCLA/2019/118.

**Supporting Information for**

**Semiconducting Metal Oxide Photonic Crystal Plasmonic Photocatalysts**

Gillian Collins[1,2,3], Alex Lonergan[1], David McNulty[1], Colm Glynn[1], Darragh Buckley[1], Changyu Hu[1], and Colm O'Dwyer[1,2,3,4]

[1] School of Chemistry, University College Cork, Cork, T12 YN60, Ireland

[2] Micro-Nano Systems Centre, Tyndall National Institute, Lee Maltings, Cork, T12 R5CP, Ireland

[3] AMBER@CRANN, Trinity College Dublin, Dublin 2, Ireland

[4] Environmental Research Institute, University College Cork, Lee Road, Cork T23 XE10, Ireland

**Additional materials characterisation**

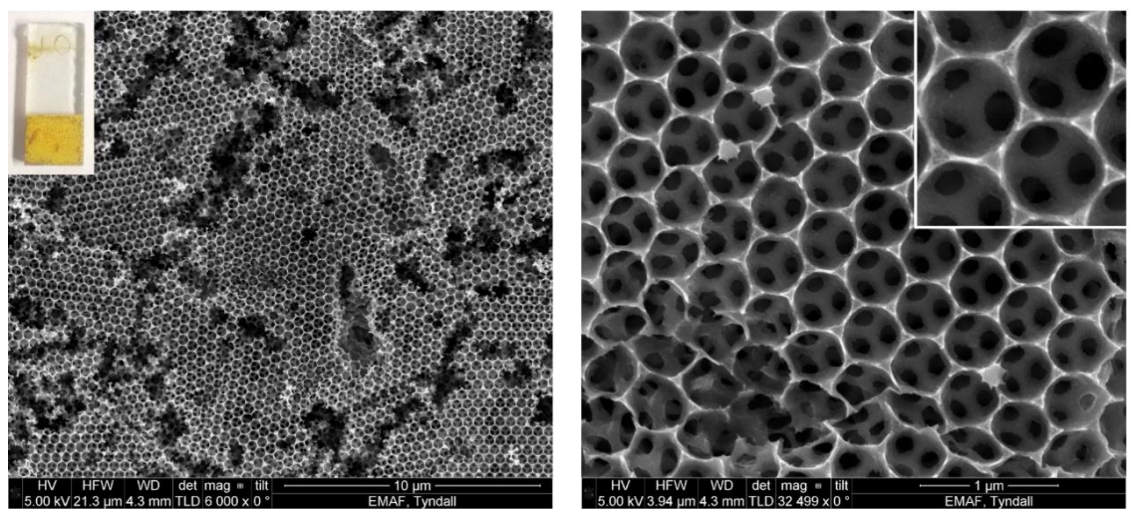

**Figure S1.** Scanning electron microscopy images of a typical $V_2O_5$ IO. Inset is a photograph of an IO film on FTO-coated glass.

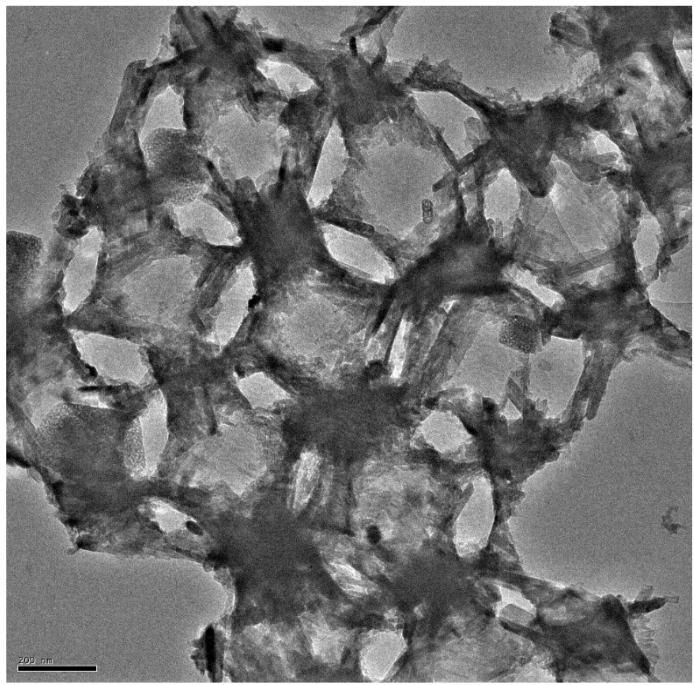

**Figure S2.** Transmission electron microscopy image of a section of $V_2O_5$ IO. Tetrahedral and octahedral infilled regions, and the pore walls that define the IO structure in two layer are visible in projection. As $V_2O_5$ is an orthorhombic crystal structure with a layered morphology, the walls are characteristically rougher compared to $TiO_2$, which maintains a nanoparticulate wall texture.

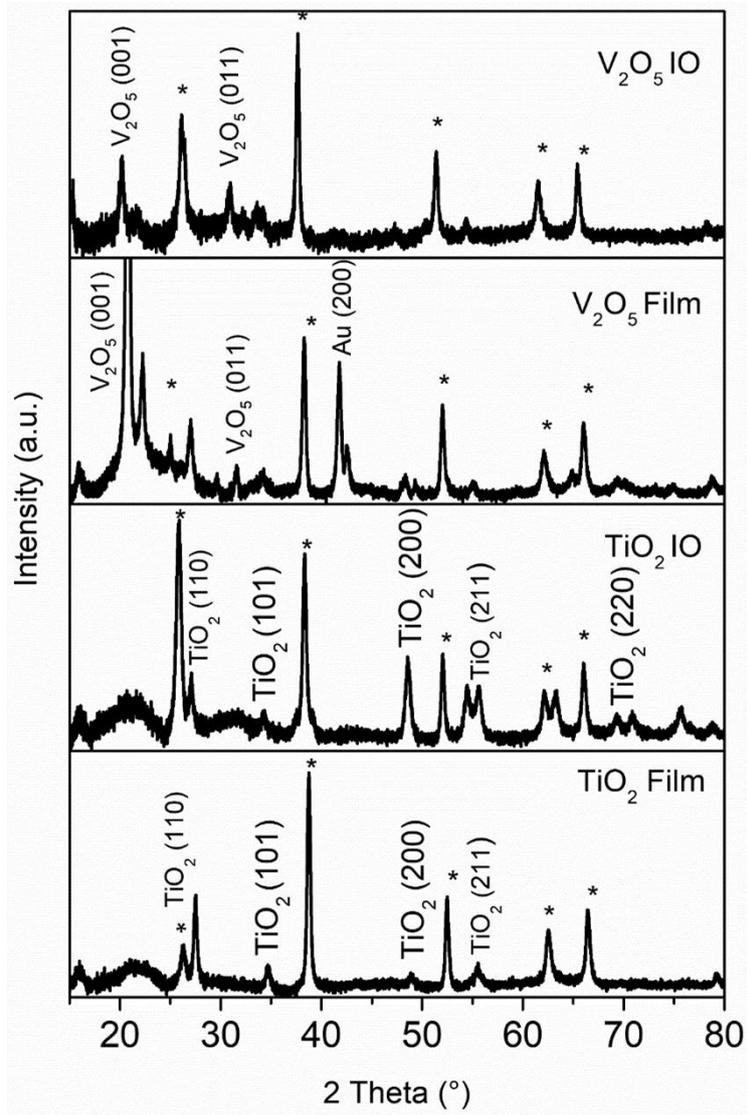

**Figure S3.** XRD patterns of $V_2O_5$ and $TiO_2$ IO and non-porous thin films on FTO. Reflections (*) are from the underlying FTO-coated glass substrate. The Au(200) reflection originates from the semi-transparent gold coating sample preparation used for SEM imaging of the thin film.

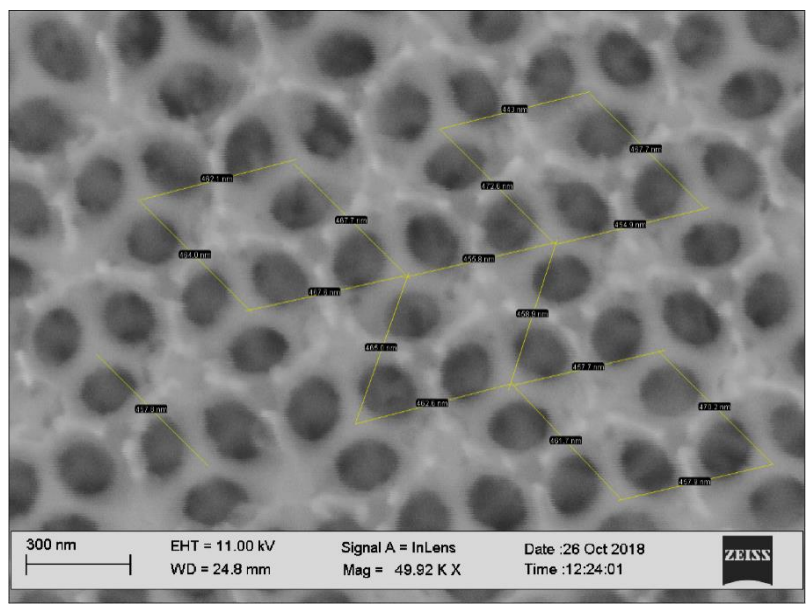

**Figure S4.** Scanning electron microscopy image a typical V₂O₅ IO. As the IO is formed from an opal template using 500 nm diameter PS spheres, the resulting IO periodicity ranges from 450-470 nm.

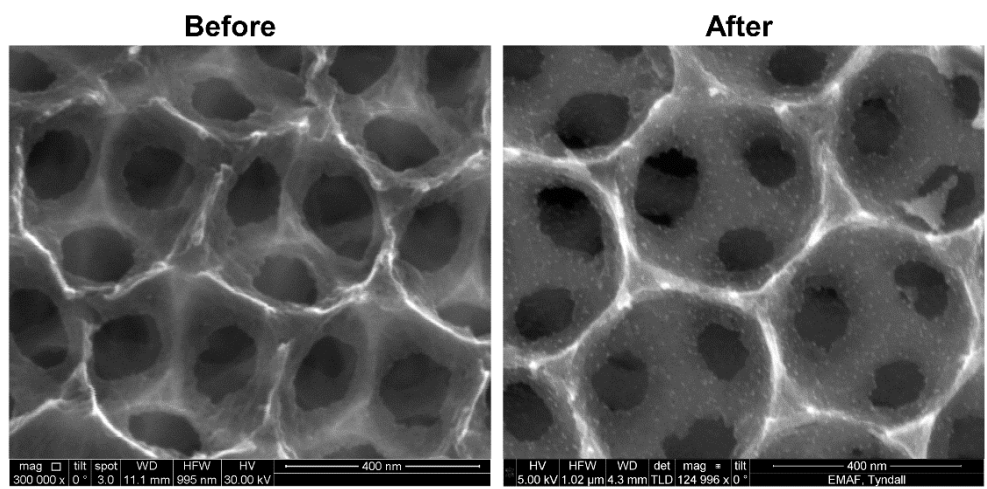

**Figure S5.** Scanning electron microscopy images of a V₂O₅ IO before and after Au NP surface functionalization.

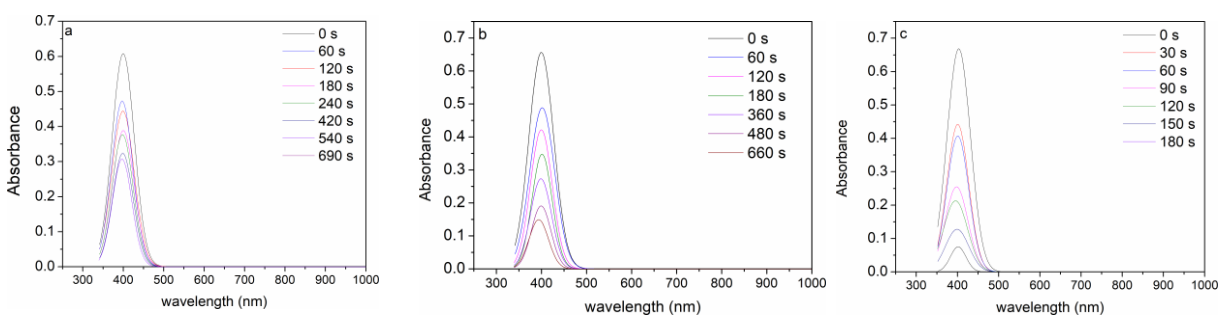

**Figure S6.** UV-vis absorbance spectra of the photocatalytic reduction of 4-nitrophenol (4-NP) (a) in the dark, (b) under broadband white light illumination and (c) illuminated with green laser light (λ = 532 nm).

## Light transmission through inverse opal photonic crystals

For straight through optical transmission, spectra were acquired using broadband white light and intensity background from FTO-coated glass was subtracted. In addition, angle-resolved measurements of TiO₂ and V₂O₅ IO shown in Fig. S7 were acquired to determine the consistency of the pseudo photonic bandgap. The measurements confirmed the stop band and transmission dip varied with angle as expected from Bragg-Snell law and associated dispersion relations.

From the Bragg-Snell relation, the transmission minimum is usually defined by $\lambda_{hkl} = \frac{2d_{hkl}}{m}\sqrt{n_{avg}^2 - sin^2\theta}$ for an FCC lattice, and angle resolved dependence can be described according to $\lambda_{hkl} = \frac{2d_{hkl}}{m}\sqrt{n_{avg}^2 - sin^2\theta}$ for the IO with air-filled pores. The transmission spectra are shown in Fig. S7 and the Bragg-Snell plot of the shift in transmission minimum (PBG) with angle is also shown.

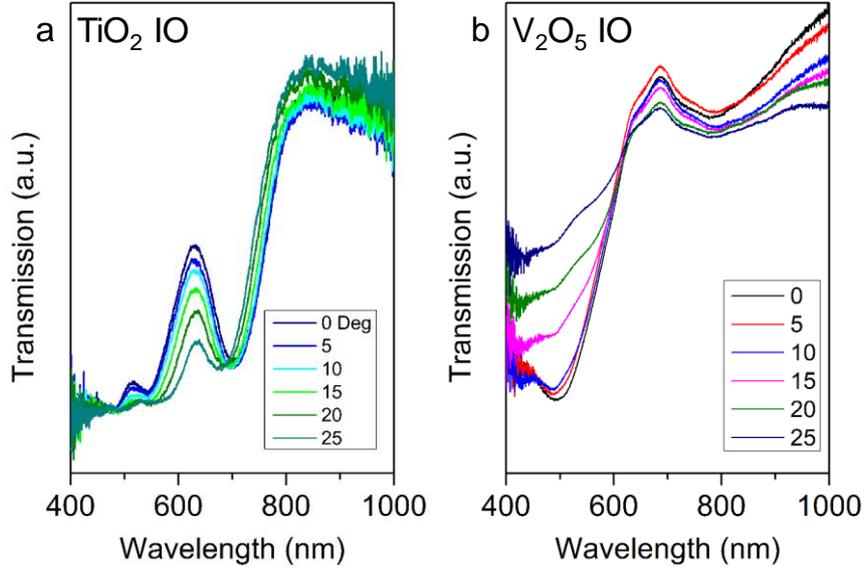

**Figure S7.** Angle-resolved optical transmission spectra for (a) TiO$_2$ IOs fashioned from infilling of an opal comprising 500 nm PS spheres, and (b) the corresponding V$_2$O$_5$ IO.

In these calculations, we use our recent modification to the standard theory for the angle-resolved spectral analysis of periodic inverse opal structures, namely that the dimension $d_{hkl} = \frac{1}{\sqrt{3}} D$. This arises from consideration of the geometric structure that interact with incident light, defining a periodic inverse opal unit cells formed from interstitial tetrahedral sites with reduced unit cell parameter of length $D$ and from interstitial octahedral sites with standard unit cell parameter of length $\sqrt{2}\, D$. Thus, the transmission dip is the spectra of TiO$_2$ IO and V$_2$O$_5$ IO are consistent with the pseudo PBG defined with solvent-filled pores according to $\lambda_{hkl}^2 = 4d_{hkl}^2 n_{avg}^2 - 4d_{hkl}^2 n_{sol}^2 \frac{n_{avg}^2}{n_{sub}^2} \sin^2\theta$, and we can thus define it's spectral location compared to the materials electronic band edge and that LSPR of the surface-immobilized Au NPs.

**Finite difference time domain model details**

Periodical IO structures can be characterized as discrete periodic changes of $\varepsilon(r)$. The method acts on photons which is analogous to the way an atomic crystalline potential acts on electrons: the discrete periodicity of $\varepsilon(r)$ in one direction that results in a dependence of $H(r)$ in that direction which is a plane wave modulated with a function $u_k^{(r)}$ that has the periodicity of the lattice, to give the Bloch state:

$$H_r(k) = e^{i(k \cdot r)} \cdot u_k^{(r)}$$

where **k** is the wavevector. For each respective **k**, we compute the corresponding band index n as a series of eigenmodes with discretely spaced frequencies. The band diagram or dispersion curve can be calculated, as each $\omega_n(k)$ changes continuously with **k**. In Fig. S8, a numerical calculation schematic of IO photonic crystal is presented. The FDTD model used 2D slabs with the same FCC packing structure to model the inverse opal in the electrolyte used during photocatalytic experiments. The pores were assumed to be completed electrolyte filled and the nominal refractive indices are summarised in the table below. The periodicity of IO is set as 420 nm, and the radius of IO pore is 380 nm. The refractive index of $TiO_2$ and $V_2O_5$ are 2.6 and 1.8 in the solvent with 1.45, respectively [1].

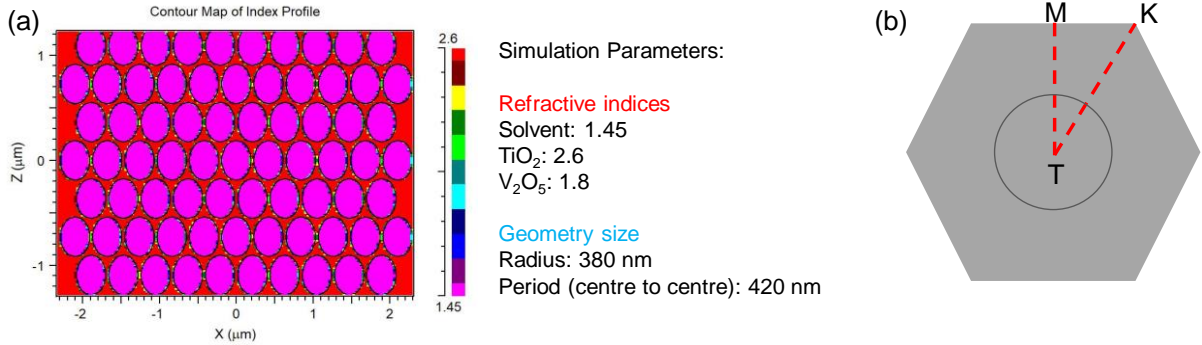

**Figure S8.** FDTD 2D simulation schematic and parameter settings are presented in (a), with a corresponding first Brillouin zone shown in (b).

Using the eigenfrequency numerical calculation, we started by calculating the eigenfrequencies at $\boldsymbol{k} = 0$, and use them as the initial inputs to iterative compute eigenmodes for small increments of $\boldsymbol{k}$. Thus, the resulting photonic band structures for $TiO_2$ and $V_2O_5$ IO are as follows:

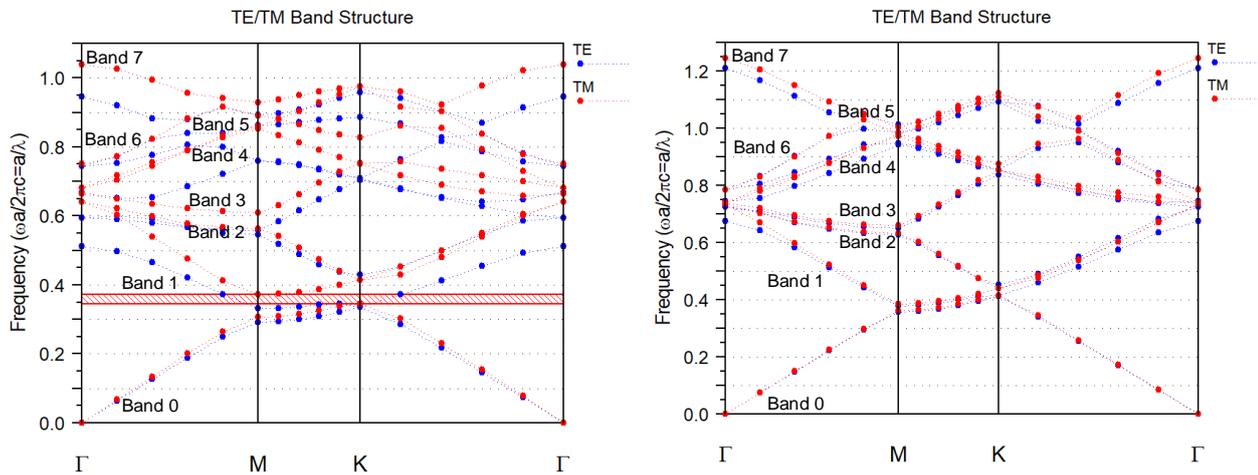

**Figure S9.** Photonic band structures of $TiO_2$ (left) and $V_2O_5$ (right) inverse opal photonic crystals by FDTD calculations in TE and TM polarizations. Red shadow region shows a band gap over the full direction.

From these, we calculated both dispersion and group index (see main text) as a function of wavelength to isolate and compare the spectra range and relative influence of slow photons caused by higher photon group velocities.

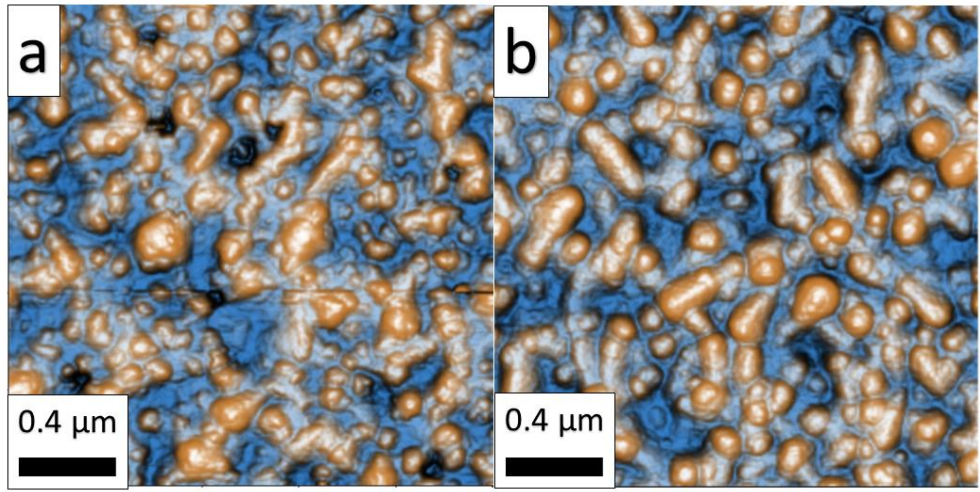

**Figure S10.** Atomic force microscopy images acquired using intermittent contact mode of the $V_2O_5$ and $TiO_2$ thin films surfaces.

[1] J. R. Devore, Refractive indices of rutile and sphalerite, J. Opt. Soc. Am. 41, 416-419 (1951)